\documentclass[8pt]{article}
\renewcommand{\baselinestretch}{1.5}
\usepackage{sectsty}
\usepackage[pdftex]{graphicx}
\usepackage{bm}
\usepackage{color}
\usepackage{textcase, url}
\usepackage[normalem]{ulem}
\usepackage[colorlinks=true, linkcolor=black, urlcolor=black, citecolor=black]{hyperref}
\usepackage{amsmath,amssymb}
\usepackage{mathtools}
\usepackage{lineno}
\usepackage{tabularx}
\usepackage{wrapfig}
\usepackage{booktabs}
\usepackage{lettrine}
\usepackage[a4paper,bottom=20mm,right=20mm,left=20mm,top=20mm]{geometry}
\usepackage[super,comma, sort&compress]{natbib}
\usepackage{float}
\usepackage{indentfirst}
\usepackage[margin=0mm]{caption}
\usepackage{subcaption}
\usepackage{siunitx}
\DeclareSIUnit{\sample}{S}
\allsectionsfont{\normalfont\sffamily\bfseries}

\newcommand{\onlinecite}[1]{\hspace{-1 ex} \nocite{#1}\citenum{#1}}

\begin{document}
\setlength{\abovedisplayskip}{8pt} % space above equation
\setlength{\belowdisplayskip}{8pt} % space below equations

\begin{center}
	\textsf{\textbf{\Large Heat-Driven Electron-Motion in a Nanoscale Electronic Circuit}}\\\vspace{2mm}
	
    {\small
		Shintaro Takada$^{1,\star}$,
        Giorgos Georgiou$^{2, 3}$,
        Everton Arrighi$^{2, 4}$,\\
        Hermann Edlbauer$^{2}$,
        Yuma Okazaki$^{1}$,
        Shuji Nakamura$^{1}$,
        Arne Ludwig$^{5}$,
        Andreas D. Wieck$^{5}$,\\
        Michihisa Yamamoto$^{6}$,
        Christopher B\"auerle$^{2}$ \&
        Nobu-Hisa Kaneko$^{1}$
	}
\end{center}
\vspace{-2mm}
\def\einr{2mm}
\def\spazi{-1.7mm}
{\small
\begin{nolinenumbers}
	\-\hspace{\einr}$^1$National Institute of Advanced Industrial Science and Technology (AIST), National Metrology Institute of Japan\vspace{\spazi}\\
	\-\hspace{\einr}\-\hspace{3mm} (NMIJ), 1-1-1 Umezono, Tsukuba, Ibaraki 305-8563, Japan\\
	\-\hspace{\einr}$^2$Univ. Grenoble Alpes, CNRS, Grenoble INP, Institut N\'{e}el, 38000 Grenoble, France\\
	\-\hspace{\einr}$^3$James Watt School of Engineering, Electronics and Nanoscale Engineering,\vspace{\spazi}\\
	\-\hspace{\einr}\-\hspace{3mm} University of Glasgow, Glasgow, G12 8QQ, United Kingdom\\
	\-\hspace{\einr}$^4$ Universit\'{e} Paris-Saclay, CNRS, Centre de Nanosciences et de Nanotechnologies, 91120, Palaiseau, France\\
	\-\hspace{\einr}$^5$ Lehrstuhl f\"{u}r Angewandte Festk\"{o}rperphysik, Ruhr-Universit\"{a}t Bochum\vspace{\spazi}\\
	\-\hspace{\einr}\-\hspace{3mm}Universit\"{a}tsstra\ss e 150, 44780 Bochum, Germany\\
	\-\hspace{\einr}$^6$Center for Emergent Matter Science, RIKEN, 2-1 Hirosawa, Wako, Saitama 351-0198, Japan\\
	\-\hspace{\einr}$^\star$ corresponding author:
	\href{mailto:shintaro.takada@aist.go.jp}{shintaro.takada@aist.go.jp}\\%\vspace{2mm}\\
\end{nolinenumbers}
}
\vspace{10mm}\vspace{-5mm}\rule{\textwidth}{0.4pt}\vspace{-6mm}
\subsection*{Abstract} 
{\bf\vspace{-3mm}
We study the interaction between two closely spaced but electrically isolated quasi-one-dimensional electrical wires by a drag experiment.
In this work we experimentally demonstrate the generation of current in an unbiased (drag) wire, which results from the interactions with a neighboring biased (drive) wire.
The direction of the drag current depends on the length of the one-dimensional wire with respect to the position of the barrier in the drag wire.
When we additionally form a potential barrier in the drive wire, the direction of the drag current is determined by the relative position of the two barriers.
We interpret this behavior in terms of electron excitations by phonon-mediated interactions between the two wires in presence of the electron scattering inside the drive wire.\\
\vspace{-3mm}\rule{\textwidth}{0.4pt}\vspace{-8mm}
}
\def\baselinestretch{1.5}\selectfont

\vspace{10mm}

%\begin{linenumbers}

%======================
% Introduction
%======================
Nanoscale electronic circuits lie at the heart of emerging quantum technologies.
The miniaturisation poses however a challenge for electrical isolation owing to interaction effects at the level of elementary particles.
The precise understanding of energy and momentum transfer between electrons and phonons is thus key for nanoscale electronic-circuit design.
An original approach to study this aspect is the so-called drag experiment, \cite{Narozhny2016} where two close-by, but electrically isolated systems are considered.
One of these parts -- the drive system -- introduces excitation via an injected electric current, whereas the other so-called drag system is sensitive to measurable responses owing to energy and momentum transfer from quantum interactions.
Studying and understanding the nature of these fundamental interactions is of paramount importance for future quantum technologies.
This is because coupled quantum systems have become essential building blocks for advanced quantum circuits employing solid-state flying qubits\cite{Bauerle2018, Yamamoto2012,Bautze2014}.

In the past, drag experiments have been successfully performed to study electron-electron and electron-phonon interactions \cite{Narozhny2016}.
Various types of nanocircuits have been investigated, such as quantum wires \cite{Debray2000,Debray2001,Morimoto2003,Yamamoto2006,Laroche2011}, quantum dots \cite{Aguado2000,Onac2006} or quantum point contacts \cite{Khrapai2007}.
These experiments revealed physical phenomena ranging from Wigner crystallisation of electrons over high-frequency-noise detection to Tomonaga-Luttinger liquids.
An interesting behaviour was observed in two adjacent but electrically isolated quantum point contacts (QPCs) that are nearly pinched\cite{Khrapai2007}.
When one of the QPCs is biased with a voltage of about \SI{1}{mV} or larger, a current with opposite direction is created in the other, unbiased QPC.
It was suggested that this counterflow is a direct consequence of an augmented thermopower effect that manifests at half conductance plateau of a QPC\cite{Molenkamp1992, Dzurak1993} as well as the asymmetric phonon-induced excitation of electrons between the two reservoirs of the drag QPC.
However, a detailed understanding of the process has not been obtained yet.

In this work, we study phonon-mediated interactions in a pair of neighboring quasi-one-dimensional (1D) wires, which are electrically isolated and equipped with potential barriers at different positions.
The placement of potential barriers at different positions allows the systematic investigation of the phonon-induced electron-motion for different configurations.
Our results corroborate the interpretation of previous experimental studies\cite{Khrapai2007} and highlight the importance of geometry for the direction of the phonon-induced current.
Our measurement data furthermore indicate that heat-driven electron-motion is strongly affected by electron-scattering within the drive wire.

\begin{figure}[htbp]
	\centerline{\includegraphics[width=0.8\textwidth]{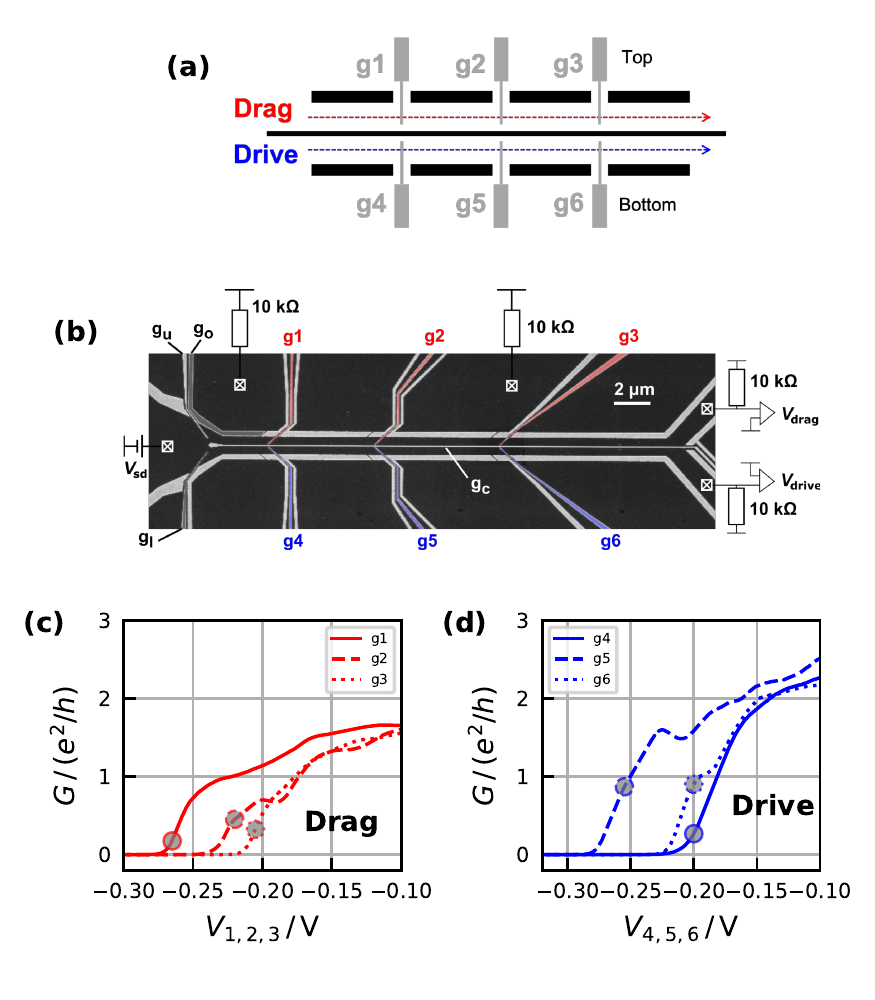}}
	\caption{
	(a) Schematic of the device for the drag experiment. The drag- (top) / drive- (bottom) wire is indicated by a red/ blue dashed arrow. A potential barrier can be selectively formed in the drag- / the drive-wire by applying voltages on the gates g1, g2, g3/ g4, g5, g6. (b) False colored SEM image of a relevant device and schematic of the measurement setup. White boxes with white crosses represent the Ohmic contacts. To perform current measurements, some of them are grounded at the base temperature through a \SI{10}{k\ohm} resistor. The Schottky gates not used in the drag experiment ($g_{\rm l}$, $g_{\rm o}$) are darkened. (c, d) Conductance across a potential barrier (c) in the drag wire (gates: g1, g2 and g3) and (d) in the drive wire (gates: g4, g5 and g6) as a function of the respective gate voltage. For the characterization of the barriers in the drag wire, the setup is changed from the basic one shown in (b). We depolarise ${\rm g_u}$ and polarise ${\rm g_l}$, ${\rm g_o}$ to directly inject the current from the left-injection contact. The circles inside the figures indicate the voltage value used for the differential drag-conductance measurements.
	}
	\label{fig:device}
\end{figure}

The investigated sample is fabricated in a GaAs/AlGaAs heterostructure hosting a two-dimensional electron gas (2DEG) at a depth of \SI{146}{nm} with electron density of \SI{1.9e11}{cm^{-2}} and mobility of \SI{1.8e6}{cm^{2}\per Vs} at \SI{4}{K}.
We perform the measurement in a dilution refrigerator at a base temperature of $\sim$ \SI{15}{mK} under zero magnetic field.
Figure\,\ref{fig:device}b shows a SEM image of the surface gates defining the investigated pair of quantum wires.
Applying a set of negative voltages on these Schottky gates, we define the potential landscape within the 2DEG and thus the electronic nanocircuit.
Surface gates that remain unused in the present experiment ($g_{\rm l}$, $g_{\rm o}$) are darkened in the SEM image (Fig.\,\ref{fig:device}b).
All gates were polarised with \SI{+0.3}{V} during the cool down to reduce the operation voltage and improve the stability of the device.
We electrostatically isolate the drag- and drive-wire (illustrated by red; top and blue; bottom in Fig.\,\ref{fig:device}a) by polarising the horizontal barrier gate (${\rm g_c}$) and the upper entrance gate (${\rm g_u}$) with a sufficiently negative voltage.
In the following we investigate the currents along the drive $I_{\rm drive}$ and drag $I_{\rm drag}$ wires as function of the voltage bias $V_{\rm sd}$ applied on the drive wire for different arrangements of potential barrier.
In these experimental scenarios, we control the presence of a potential barrier via a negative voltage applied on the corresponding Schottky gates g1 - g3 in the (top) drag or g4 - g6 in the (bottom) drive wire \cite{pbarrier}.
For the measurements, the side gates of the two wires are tuned to host $4 \sim 5$ conduction modes in each wire in order to suppress significant electron-scattering due to disorders along the transport paths.
To suppress the influence of thermal voltage variation between different measurement lines, all Ohmic contacts, except the injection contact on the left, are connected to ground at the base temperature through a \SI{10}{k\ohm} resistor.
The currents flowing through the drive and drag wire are respectively obtained by measuring the voltages across the \SI{10}{k\ohm} resistor.
If not indicated otherwise, the drag current is obtained by measuring the voltage $V_{\rm drag}$ at the right-most Ohmic contact as shown in Fig 1b.

To begin, we characterise the potential barriers.
Figures\,\ref{fig:device}c (\ref{fig:device}d) show conductance measurements as a function of the voltage applied on the barrier gates g1 -- g3 (g4 -- g6) along the drag (drive) wire.
The data shows plateau-like features near the conductance pinch-off\cite{note1} that are consistent for both drive and drag wires.
By selectively polarising the barrier gates, we form a potential barrier at deliberately chosen positions along the drag and drive wires.

\begin{figure}[htbp]
	\centerline{\includegraphics[width=0.8\textwidth]{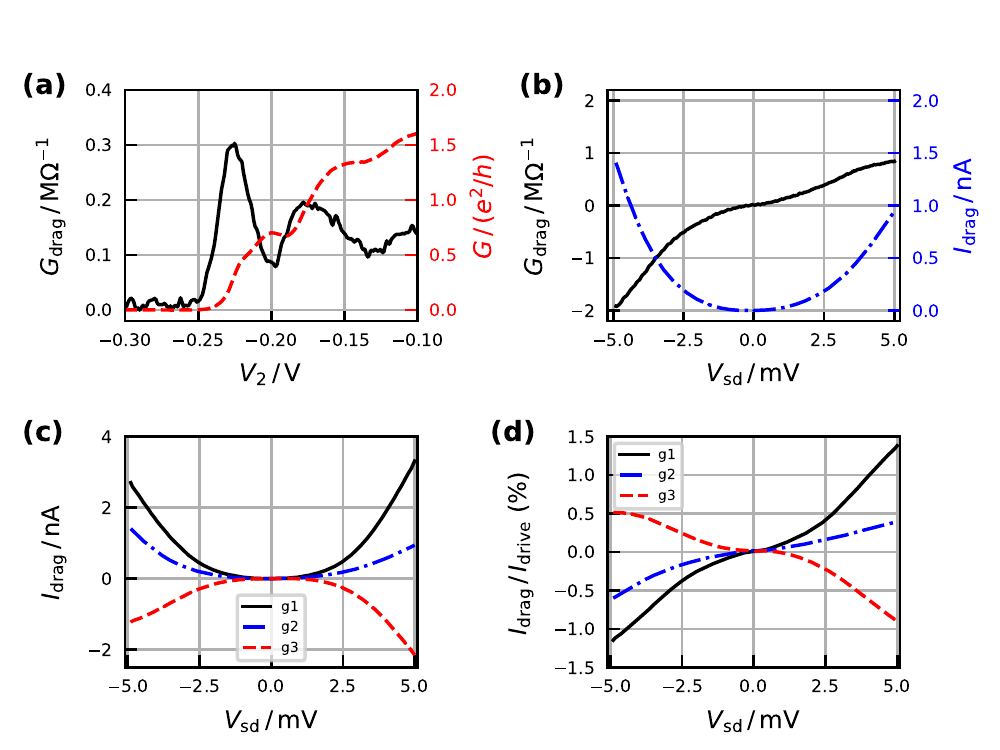}}
	\caption{
		(a)  Differential drag conductance (black solid curve, left axis) and conductance (red dashed curve, right axis) as a function of the voltage ($V_{\rm 2}$) on gate g2. The conductance is measured in a separate measurement and these are the same data as the red dashed curve in Fig.\,\ref{fig:device}c. For the differential drag conductance measurement we fix $V_{\rm sd}$ to \SI{2}{mV}. (b) Differential drag conductance (black solid curve, left axis) and drag current (blue dash-dotted curve, right axis) as a function of the dc bias $V_{\rm sd}$ when the potential barrier at g2 is fixed to the gate voltage indicated by the red dashed circle in Fig.\,\ref{fig:device}c. The drag current is obtained by integrating the differential drag conductance along the bias voltage across the drive wire \cite{measurement}. (c) Drag current as a function of $V_{\rm sd}$ when we form a potential barrier at 3 different positions in the drag wire. For the 3 curves, the voltage on the potential barrier is fixed to the position indicated by the circles in Fig.\,\ref{fig:device}c. (d) The data of (c) is plotted in the form of the drag current over the drive current as a function of $V_{\rm sd}$.
	}
	\label{fig:ub}
\end{figure}

The measurements are performed in the basic setup illustrated in Fig.\,\ref{fig:device}a.
Here we form a potential barrier only in the drag wire and the gates g4, g5, and g6 in the drive wire are not used.
To measure the differential drag conductance $G_{\rm drag}$, we inject current into the drive wire by applying an AC voltage drive of $V^{\rm rms} =$ \SI{70.7}{\micro V} at the left Ohmic contact and at a frequency of \SI{23.3}{Hz}, which is used for lock-in detection purposes.
In addition to the AC voltage drive, we apply a DC offset bias, $V_{\rm sd}$, on the injection contact and measure the resulting induced AC current in the drag wire to obtain $G_{\rm drag}$ at a fixed $V_{\rm sd}$ \cite{measurement}.
In this experiment the polarity of $G_{\rm drag}$ gives the relation between the flow-direction of the drive current and the one of the drag current.
When $G_{\rm drag}$ is positive, the drive current and the drag current flow in the same direction.
On the other hand, when $G_{\rm drag}$ is negative, the drive current and the drag current flow towards opposite directions.
The black solid curve in Fig.\,\ref{fig:ub}a shows the differential drag conductance, $G_{\rm drag}$, as a function of the voltage applied on the potential barrier gate g2 defined by $V_2$, while the offset bias $V_{\rm sd}$ was fixed to \SI{2}{mV}.
For comparison purposes, the red dashed curve illustrates the corresponding conductance $G$ of the same potential barrier in the absence of the DC bias voltage.
Clear peaks of $G_{\rm drag}$ at the voltage around half of the conductance plateaus are visible.
At those peaks the thermopower of the potential barrier becomes large \cite{Molenkamp1992, Dzurak1993} and hence a large drag current flows when a temperature-gradient is induced across the potential barrier.
Here we confirm that as far as we keep a few conduction modes within each wire, a slight variation of the drive-wire-width will not change the gate voltage dependence of the differential drag conductance qualitatively \cite{wire_width}.
We then fix the potential barrier, $V_2$, at the voltage where we obtain a peak in the differential drag conductance $G_{\rm drag}$ as shown by the dashed circle in Fig.\,\ref{fig:device}c and remeasure $G_{\rm drag}$ as a function of the DC bias voltage, $V_{\rm sd}$.
The result is shown by the black solid curve in Fig.\,\ref{fig:ub}b, in which $G_{\rm drag}$ appears to be positive for positive $V_{\rm sd}$ and negative for negative $V_{\rm sd}$.
From $G_{\rm drag}$ as a function of $V_{\rm sd}$ we calculate the drag current $I_{\rm drag}$ \cite{measurement} (blue dash-dotted curve in Fig.\,\ref{fig:ub}b).
We clearly observe that the direction of the drag current does not depend on the direction of the drive current (i.e. the polarity of $V_{\rm sd}$).
This result indicates that the temperature-gradient across the potential barrier depends on the absolute value of the drive current.
Here electrons at the right side of the potential barrier are excited more than the ones at the left side.
As a result, electrons flow from the right to the left in the drag wire.

We also perform a similar measurement by forming a potential barrier either at g1 or g3.
As for the case of g2, the differential drag conductance, $G_{\rm drag}$, shows a peak for the potential barrier at g1, however it shows a dip for the barrier formed at g3.
After fixing the voltage applied to the potential barriers, $V_{\rm 1}$ or $V_{\rm 3}$, at their peak/dip value respectively (indicated by the circles in Fig.\,\ref{fig:device}c) we perform a DC bias voltage scan and calculate the drag current.
The results are summarised in Fig.\,\ref{fig:ub}c, including the data from g2 shown in Fig.\,\ref{fig:ub}b for comparison.
For g1 the drag current is positive as for the case of g2 while for g3 it is negative.
When we assume a homogeneous energy exchange between the drag and the drive wire, the relative length of the drag wire with respect to the position of the potential barrier in the drag wire determines on which side of the potential barrier electrons are more excited.
For g1 and g2, the wire at the right side of the potential barrier is longer while for g3 the wire at the left side is longer.
This explains the observed behaviour of the drag current.
We plot the data of Fig.\,\ref{fig:ub}c in Fig.\,\ref{fig:ub}d in the form of $I_{\rm drag}/I_{\rm drive}$ to clearly present the strength of the drag effect.
The observed drag effect is orders of magnitude stronger than the one expected from the Coulomb drag \cite{Raichev2000} in our well-separated wires (the distance between the drag and the drive wire is $\sim$ \SI{300}{nm}).
This suggests that phonon-mediated interactions are at the origin of energy exchange between the two wires.
Here when we carefully look at the data around zero DC bias, $I_{\rm drag}/I_{\rm drive}$ starts appearing at a few hundred \si{\micro V}.
This energy scale corresponds to the subband energy gap in the drive wire, which is also visible as a small dip structure in $I_{\rm drive}$ (See section \ref{suppl:idrive} of Supplementary material\cite{measurement}).
For $V_{\rm sd}$ above a few hundred \si{\micro V}, the inter-subband electron-scattering becomes possible and hence scattering with a large momentum transfer occurs \cite{DeGottardi2019}.
Such a process should play a key role for the observed drag effect and may enhance the electron-phonon coupling.

\begin{figure}[htbp]
	\centerline{\includegraphics[width=0.8\textwidth]{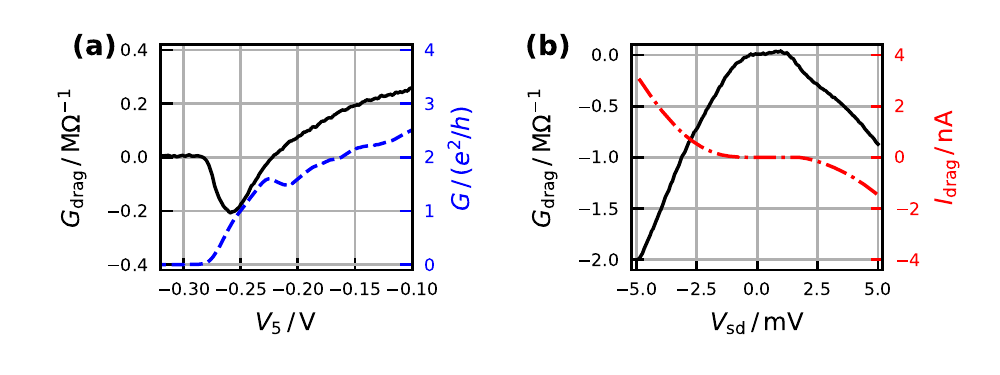}}
	\caption{
		(a) Differential drag conductance (black solid curve, left axis) and conductance (blue dashed curve, right axis) as a function of the voltage ($V_5$) on gate g5. The conductance is measured in a separate measurement and these are the same data as the blue dashed curve in Fig.\,\ref{fig:device}d. For the differential drag conductance measurement we fix $V_{\rm sd}$ to \SI{2}{mV}. Here a potential barrier is formed at g2 in the drag wire by setting $V_2$ to the value indicated by the red dashed circle in Fig.\,\ref{fig:device}c. (b) Differential drag conductance (black solid curve, left axis) and drag current (red dash-dotted curve, right axis) as a function of the dc bias $V_{\rm sd}$ when $V_5$ is fixed close to its dip value shown in (a) (the value is also indicated in Fig.\,\ref{fig:device}d by the blue dashed circle). The drag current is obtained by integrating the differential drag conductance along the bias voltage across the drive wire \cite{measurement}.
	}
	\label{fig:lb}
\end{figure}

In what follows, we investigate how the drag current is affected by introducing potential barriers along the drive wire.
To do this we return back to the basic setup (Fig.\,\ref{fig:device}a) and form a potential barrier along the drag wire at gate g2 by setting $V_2$ at the voltage value illustrated by the dashed circle shown in Fig.\,\ref{fig:device}c.
Then we sweep the voltage $V_5$ on gate g5 and measure the differential drag conductance $G_{\rm drag}$ at $V_{\rm sd} = $\SI{2}{mV}.
The result is shown in Fig.\,\ref{fig:lb}a.
The differential drag conductance, $G_{\rm drag}$ (black solid curve in Fig.\,\ref{fig:lb}a), shows a positive value when $V_5$ is less negative and thus no potential barrier is formed in the drive wire.
This is consistent with the result shown in Fig.\,\ref{fig:ub}b.
When $V_5$ is decreased even further, $G_{\rm drag}$ crosses zero and a negative dip appears before the pinch off.
To study the origin of this dip, we fix $V_5$ to that value (indicated by the dashed circle in Fig.\,\ref{fig:device}d) and measure $G_{\rm drag}$ as a function of input bias voltage $V_{\rm sd}$.
The results shown in Fig.\,\ref{fig:lb}b demonstrate that the drag conductance is negative for both positive and negative $V_{\rm sd}$.
This means that the drag current and the drive current flow in opposite direction.
Such a counterflow of electrons has been reported in Ref.\,\onlinecite{Khrapai2007}.
In this reference the asymmetric phonon-induced excitation of electrons is suggested to explain the counterflow behaviour.
When the potential barrier in the drive wire is set close to the pinch off, the tunnel-probability across the barrier becomes energy dependent and electrons with higher energy are transmitted more favorably.
As a consequence energy relaxation occurs mainly after the barrier by the emission of phonons.
This results in a counterflow of the electrons in the drag wire.
We confirm that the origin of the induced drag current is indeed an asymmetric excitation with respect to the potential barrier \cite{asymmetry}.
Although the discussed counterflow behaviour of the drag current is qualitatively same for our work and Ref.\,\onlinecite{Khrapai2007}, the observed drag current is much larger in our work ($\sim$ \si{nA} while $\sim$ \si{pA} in Ref.\,\onlinecite{Khrapai2007}).
The difference is considered to originate from the potential barrier being embedded in a quasi 1D wire in our work, while in Ref.\,\onlinecite{Khrapai2007} it was embedded in a 2D reservoir.

\begin{figure}[htbp]
	\centerline{\includegraphics[width=0.8\textwidth]{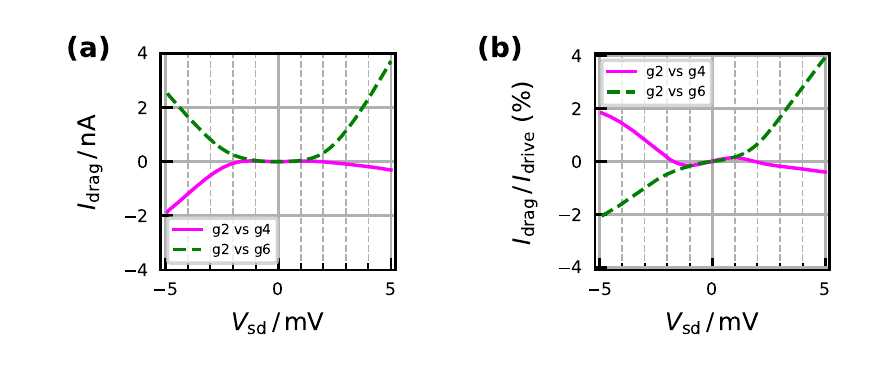}}
	\caption{
		(a) Drag current as a function of $V_{\rm sd}$ measured by forming a potential barrier at different positions in both drag and drive wires. Here a potential barrier in the drag wire is formed at g2 by setting $V_{\rm 2}$ to the value indicated by the red dashed circle in Fig.\,\ref{fig:device}c. In the drive wire a potential barrier is formed either at g4 (magenta solid curve) or at g6 (green dashed curve). The voltage on the gates g4 and g6 are set to the values indicated by the blue solid and dotted circles shown in Fig.\,\ref{fig:device}d, respectively. (b) The data of (a) is plotted in the form of the drag current over the drive current as a function of $V_{\rm sd}$.
	}
	\label{fig:lb3}
\end{figure}

Finally we also investigate the drag current by forming a potential barrier at different positions in each wire.
For this measurement we form a potential barrier at gates g2 and g4 by setting their voltages to their respective values indicated by the dashed and solid circles in Fig.\,\ref{fig:device}c and Fig.\,\ref{fig:device}d, respectively.
In this situation we observe a negative drag current for both positive and negative $V_{\rm sd}$ as shown by the magenta solid curve in Fig.\,\ref{fig:lb3}a.
When we perform the same measurement by placing the potential barrier at g6 instead of g4, the direction of the drag current becomes positive for both signs of $V_{\rm sd}$ as shown by the green dashed curve in Fig.\,\ref{fig:lb3}a.
This result is in clear contrast to the result reported in Ref.\,\onlinecite{Khrapai2007}, where the counterflow behaviour is preserved in a similar drag-current measurement performed for the QPC potential barriers separated by $\sim$ \SI{300}{nm}.
In our device the distance between the potential barriers is much larger ($l_{\rm g2-g4} =$ \SI{6}{\micro m}, $l_{\rm g2-g6} =$ \SI{7}{\micro m}).
Here the drag current is generated by electrons, which absorb phonons with high enough energy and can pass through the potential barrier in the drag wire.
Our results indicate that phonon-mediated energy transfer between the two wires contributing to the drag current mainly occurs within a distance of up to \SI{6}{\micro m}.
The flight time of electrons over \SI{6}{\micro m} is about \SI{30}{ps}.
The electron-phonon scattering time $\tau_{e-p}$ reported for GaAs-based 2DEG is roughly \SI{1}{ns}\cite{Verevkin1996}, which is much larger than the above-mentioned flight time.
Our results suggest that most of the electrons, which make it over the potential barrier in the drive wire and can excite transport across the barrier in the drag wire, lose their excess energy within a few micrometers before their energy is transferred to phonons in our 1D wire.
The decay of energy and momentum most probably occurs through inter-subband electron-electron scattering \cite{DeGottardi2019}.
This is in line with previously reported electron-electron scattering times $\tau_{\rm e-e}$ in 1D wires or 2DEG with values ranging on the order of \SI{10}{ps} or shorter for high-energy electrons ($V_{\rm sd} > $\SI{1}{mV}) \cite{Giuliani1982, wind1986, Tikhonov2014}.
In Fig.\,\ref{fig:lb3}b we plot the data of Fig.\,\ref{fig:lb3}a in the form of $I_{\rm drag}/I_{\rm drive}$, which is an indicator of the strength of the drag effect and hence provides evidence on the strength of electron-phonon scattering.
For small bias ($|V_{\rm sd}| <$ \SI{1}{mV}) both curves show similar tendency and $I_{\rm drag}/I_{\rm drive}$ increases with a small positive slope as a function of $V_{\rm sd}$.
This behaviour is similar to the results when the potential barrier is absent in the drive wire as shown in the blue dash-dotted curve in Fig.\,\ref{fig:ub}d.
For the low bias regime ($|V_{\rm sd}| <$ \SI{1}{mV}) the drag current is induced by the thermal excitation over the long drive wire, where inter-subband electron-electron scattering is still possible with this bias.
On the other hand, for the high bias regime ($|V_{\rm sd}| >$ \SI{1}{mV}) the asymmetric, phonon-induced heating by high energy electrons transmitting across a potential barrier seem to dominate the drag current as discussed in Figs.\,\ref{fig:lb} and \ref{fig:lb3}a.
The two curves show opposite slopes as a function of $V_{\rm sd}$, which is consistent with the opposite signs observed for the drag currents in Fig.\,\ref{fig:lb3}a.
For $|V_{\rm sd}| >$ \SI{2}{mV}, where inter-subband scattering becomes possible around the barrier \cite{note2}, $I_{\rm drag}/I_{\rm drive}$ increases significantly with $V_{\rm sd}$.
This indicates that electron-phonon scattering is locally enhanced there.
Finally, $I_{\rm drag}/I_{\rm drive}$ observed for these conditions is about $1\, \%$.
This small current ratio is consistent with the scenario in which most of the electrons in the drive wire do not emit phonons with high enough energy to contribute to the drag current due to the electron-electron scattering.
It is also consistent with the ratio between the flight time of electrons over a few \si{\micro m} and the electron-phonon scattering time.

In summary, a drag-type measurement in two adjacent but electrically isolated one-dimensional quantum wires has been performed.
Our results demonstrate that phonon-mediated energy transfer has an important effect in one-dimensional nanocircuits and can induce electron-flow in an adjacent, but electrically isolated wire when a potential barrier is formed.
The direction of the induced electron-flow is parallel or anti-parallel to the drive current depending on the respective position of the potential barriers in the nanocircuit.
Furthermore, our results indicate that phonon-emission from electrons transmitted through a potential barrier occurs within a short distance ($<$ \SI{6}{\micro m}) in the non-linear regime ($V_{\rm sd} >$ \SI{1}{mV}).
This suggests that momentum relaxation occurs before the energy relaxation by electron-phonon interaction.
Since a potential barrier is one of the key elements in quantum electronic circuits \textemdash \, a prime example being a beam splitter \textemdash \, our results will provide useful information for quantum operations in nanocircuits in particular for quantum circuits containing a potential barrier.

\vspace{30pt}
%\acknowledgment
We acknowledge fruitful discussion with Stefan Ludwig, and Vadim Khrapai. S.T. acknowledges the financial support by JSPS. Grand-in-Aid for Scientific Research B (No.20H02559). A.L. and A.D.W. acknowledge support by DFG-TRR160, DFG project 383065199, EU Horizon 2020 Grant No. 861097, BMBF - QR.X KIS6QK4001, and DFH/UFA CDFA-05-06. N.-H.K. was partly supported by JSPS KAKENHI Grant Number JP18H05258. G.G. and C.B. acknowledge financial support from the French Agence Nationale de la Recherche (ANR), project QTERA ANR-2015-CE24-0007-02 and E.A. and C.B. acknowledge financial support from the French Agence Nationale de la Recherche (ANR), project FullyQuantum ANR-16-CE30-0015. This project has received funding from the European Union’s H2020 research and innovation programme under grant agreement No 862683. This work was supported by a joint JST - ANR Research Project through Japan Science and Technology Agency, CREST (grant number JPMJCR1876) and the French Agence Nationale de la Recherche, Project QCONTROL ANR-18-JSTQ-0001.

\bibliographystyle{jpsj}

\bibliography{dcdrag}

\begin{thebibliography}{10}
\bibitem{pbarrier}
Remark on depletion below unpolarized barrier gates (g1 - g6) via the
  polarization of neighboring side gates: When applying a negative voltage on a
  Schottky gate, the potential extension exponentially decreases if screening
  effects from a neighboring gate are present. The order of magnitude of the
  decaying potential at the 2DEG plane is its depth -- that is \SI{146}{nm} in
  the presently investigated device. Here the barrier gate is sandwiched by the
  gates to form the drag and drive wires and the separation between those
  neighboring side gates is about \SI{220}{nm}. Therefore, when we form the
  drag and drive wires in the experiment by applying a sufficiently large
  negative voltage on the side gates, the 2DEG under the barrier gate
  sandwiched by the side gates is depleted also when it is unpolarized.
\bibitem{note1}
The conductance is lower than the quantised value $2e^2/h$ at a plateau-like
  feature partially due to an additional series resistance introduced by our
  semi-ballistic wire. The length of the drive and drag wire is about
  \SI{20}{\micro m}, which is longer than the elastic mean-free-path ($l_e \sim
  $\SI{13}{\micro m}) of the 2DEG employed in this experiment. Due to the
  semi-ballistic nature of our long device, electrons travelling through the
  wires experience scattering which in turn decreases the conductance. We also
  comment that the potential barrier employed in this experiment is not a QPC
  formed by split gates and hence well-quantised plateaus are not observed.
\bibitem{note2}
The subband energy spacing $\Delta$ is increased around the potential barrier
  due to the additional potential from the gate forming the potential barrier.
\end{thebibliography}

%======================
% Device
%======================
%%% Figure 1: Device and characterization of potential barriers %%%%%%%
%%%%%%%%%%%%%%%%%%%%%%%%

%%% Figure 2: Barrier in the drag wire %%%%%%%
%%%%%%%%%%%%%%%%%%%%%%%%%%%%%%%%%%%%%%%%

%%% Figure 3: Barriers in both the drag wire and the drive wire %%%%%%%
%%%%%%%%%%%%%%%%%%%%%%%%

%%% Figure 4: Drag current for the shifted potential barriers %%%%%%%
%%%%%%%%%%%%%%%%%%%%%%%%

% Supplement
\clearpage
\onecolumn
\setcounter{page}{1}

\renewcommand{\thefigure}{S\arabic{figure}}
\setcounter{figure}{0}

\begin{center}
    \textsf{\textbf{\Huge Supplemental material\\\vspace{2mm}
    \Large Heat-Driven Electron-Motion in a Nanoscale Electronic Circuit}}\\\vspace{5mm}
%\vfill
    {\small
        Shintaro Takada$^{1,\star}$,
        Giorgos Giorgiou$^{2,3}$,
        Everton Arrighi$^{2,4}$,
        Hermann Edlbauer$^{2}$,
        Yuma Okazaki$^{1}$,
        Shuji Nakamura$^{1}$,
        Arne Ludwig$^{5}$,
        Andreas D. Wieck$^{5}$,
        Michihisa Yamamoto$^{6}$,
        Christopher B\"{a}uerle$^{2}$, and
        Nobu-Hisa Kaneko$^{1}$
    }
\end{center}
\vspace{2mm}
\def\einr{2mm}
\def\spazi{-1.7mm}
{\small
    \-\hspace{\einr}$^1$ National Institute of Advanced Industrial Science and Technology (AIST), National Metrology Institute of Japan (NMIJ), Tsukuba, Ibaraki 305-8563, Japan\\
    \-\hspace{\einr}$^2$ Univ. Grenoble Alpes, CNRS, Grenoble INP, Institut N\'{e}el, 38000 Grenoble, France\\
    \-\hspace{\einr}$^3$ James Watt School of Engineering, Electronics and Nanoscale Engineering, University of Glasgow, Glasgow, G12 8QQ, United Kingdom\\
    \-\hspace{\einr}$^4$ Universit\'{e} Paris-Saclay, CNRS, Centre de Nanosciences et de Nanotechnologies, 91120, Palaiseau, France\\
    \-\hspace{\einr}$^5$ Angewandte Fesk\"{o}rperphysk, Ruhr-Universit\"{a}t Bochum, D-44780 Bochum, Germany\\
    \-\hspace{\einr}$^6$ Center for Emergent Matter Science, RIKEN, 2-1 Hirosawa, Wako, Saitama 351-0198, Japan\\
    \-\hspace{\einr}$^\star$ shintaro.takada@aist.go.jp
}
\section{Measurement of the differential drag conductance and the drag / drive current}
\label{suppl:idrive}

The measurements of the differential drag conductance are performed by applying an AC excitation (\SI{23.3}{Hz}, $V^{\rm rms} = $\SI{70.7}{\micro V}) with a controlled DC offset bias ($V_{\rm sd}$) on the left injection contact, which injects the current into the drive wire.
To detect the drive and drag currents we measure the AC voltage across the \SI{10}{k\ohm} resistor for the drive wire ($V_{\rm drive}$) and for the drag wire ($V_{\rm drag}$) respectively with a standard lock-in technique.
In this measurement setup the voltage across the drive wire ($V^{\rm rms}_{\rm drive}$) is expressed by
\begin{equation}
    V^{\rm rms}_{\rm drive} = V^{\rm rms} - V_{\rm drive } - V_{\rm drive} \cdot \frac{R_{\rm injection}+R_{\rm drive}}{10\, {\rm k\Omega}},
\end{equation}
where $R_{\rm injection}$ is the resistance of the injection contact and $R_{\rm drive}$ is the resistance of the right detection contact in the drive wire.
Using the above formula for the $V^{\rm rms}_{\rm drive}$, the differential conductance of the drive wire $G_{\rm drive}$ and differential drag conductance of the drag wire $G_{\rm drag}$ are obtained by
\begin{equation}
    G_{\rm drive / drag} = \frac{V_{\rm drive / drag}}{10\,{\rm k\Omega} \cdot V^{\rm rms}_{\rm drive}}.
\end{equation}
When a finite DC offset bias $V_{\rm sd}$ is applied on the injection contact, the DC bias across the drive wire becomes
\begin{equation}
    V_{\rm sd,\ drive} = V_{\rm sd} \cdot \frac{V^{\rm rms}_{\rm drive}}{V^{\rm rms}}.
\end{equation}
%%% Figure S1: Drive current vs V_sd %%%%%%%
\begin{figure}[htbp]
	\centerline{\includegraphics[width=0.9\textwidth]{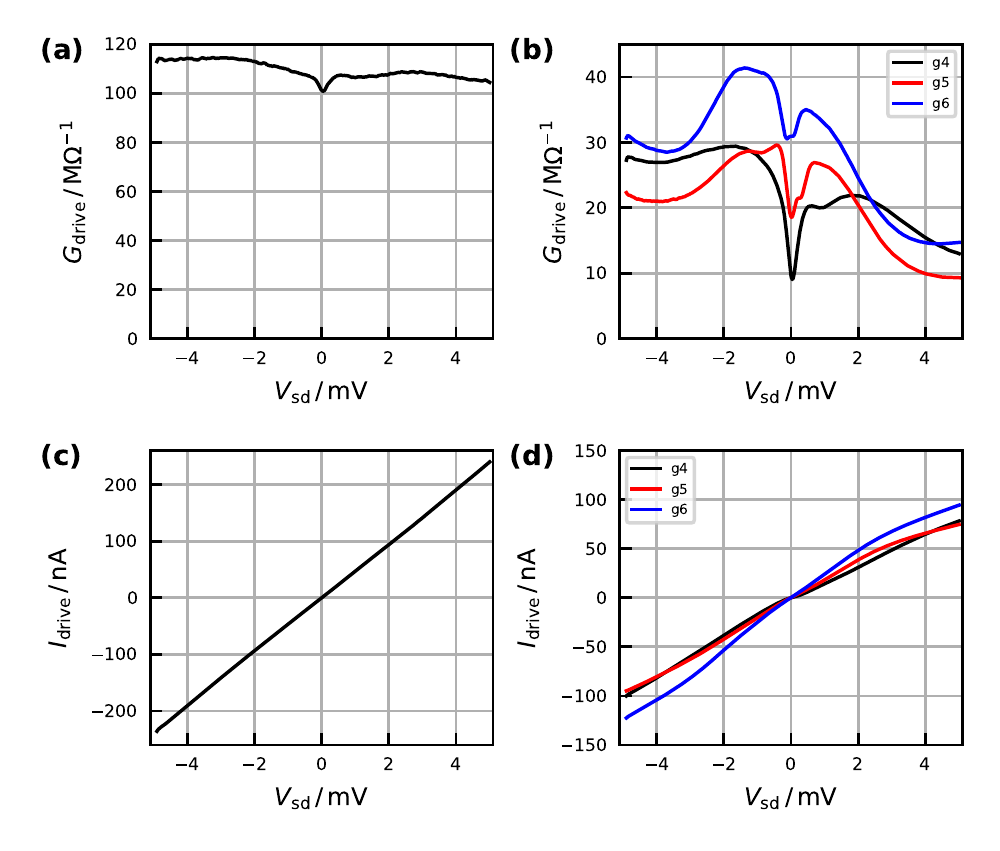}}
	\caption{
	(a) Differential conductance of the drive wire / (c) drive current as a function of $V_{\rm sd}$ when there is no potential barrier in the drive wire. (b) Differential conductance of the drive wire / (d) drive current as a function of $V_{\rm sd}$ when a potential barrier is formed at either g4 (black curve), g5 (red curve) or g6 (blue curve) in the drive wire.
	}
	\label{fig:idrive}
\end{figure}
%%%%%%%%%%%%%%%%%%%%%%%%
By measuring the differential conductance of the drive wire and the differential drag conductance of the drag wire at different values of $V_{\rm sd}$ from zero and by integrating them along $V_{\rm sd,\ drive}$, the current flows in the drive wire ($I_{\rm drive}$) and the drag current ($I_{\rm drag}$) are obtained in the experiment.

Fig.\,\ref{fig:idrive}a shows the differential conductance of the drive wire when there is no barrier in the drive wire as a function of $V_{\rm sd}$.
A small dip with a width of a few hundred \si{\micro V} appears around zero bias voltage.
This structure is considered to be associated with the subband energy gap of the drive wire and it is an order of a few hundred \si{\micro V}.
Fig.\,\ref{fig:idrive}c shows the drive current obtained from the data in Fig.\,\ref{fig:idrive}a as explained above.
The drive current increases almost linearly as a function of $V_{\rm sd}$.
Fig.\,\ref{fig:idrive}b shows the differential conductance of the drive wire when a potential barrier is formed at g4 (black curve), g5 (red curve) and g6 (blue curve), respectively.
Here the absolute value of the differential conductance is limited by the potential barrier compared to the case without a potential barrier (Fig.\,\ref{fig:idrive}a) and a larger structure with a width of $\sim$ \si{mV} appears around zero bias.
This larger structure is considered to be associated with the enhanced subband energy gap around the potential barrier.
The drive current in this situation is plotted in Fig.\,\ref{fig:idrive}d.
The drive current shows non-linear dependence on $V_{\rm sd}$ for $V_{\rm sd} \gtrsim \ $\SI{2}{mV}.
This non-linear dependence originates from the structure in $G_{\rm drive}$ discussed above and is considered to be associated with inter-subband scattering for $V_{\rm sd}$ above the subband energy gap around the potential barrier.

\newpage

\section{Gate voltage dependence of the differential drag conductance for different drive-wire widths}
Here we investigate the gate voltage dependence of the differential drag conductance $G_{\rm drag}$ at g2 for different drive-wire-widths.
The measurement configuration is basically same as the configuration employed to measure the data plotted in Fig.\,\ref{fig:ub}a.
However, the voltage on the gate ${\rm g_c}$ is set to be slightly (\SI{50}{mV}) more positive, where the two wires are still well separated but the width of the two wires become slightly wider.
For this measurement $V_{\rm sd}$ is fixed to \SI{2}{mV} as for the case in Fig.\,\ref{fig:ub}a.
To modify the width of the drive wire we change the voltage on all the lower side gates at the same time between \SI{-0.3}{V} and \SI{-0.6}{V}.
\SI{-0.3}{V} is the value used for the measurements in the main paper.
At this voltage the drive wire hosts $4 \sim 5$ conduction modes.
When the value is changed to \SI{-0.6}{V}, the number of the conduction modes is reduced to $2 \sim 3$.
For the different drive-wire-widths the absolute value of the differential drag conductance slightly changes but the main feature, in particular the peak near the conductance pinch-off, is qualitatively the same.
This result indicates that the observed drag behaviour does not change for the variation of the drive-wire-width investigated here (hosting between $2 \sim 3$ and $4 \sim 5$ conduction modes).
%%% Figure S2: Differential drag conductance for different drive wire width %%%%%%%
\begin{figure}[htbp]
	\centerline{\includegraphics[width=0.7\textwidth]{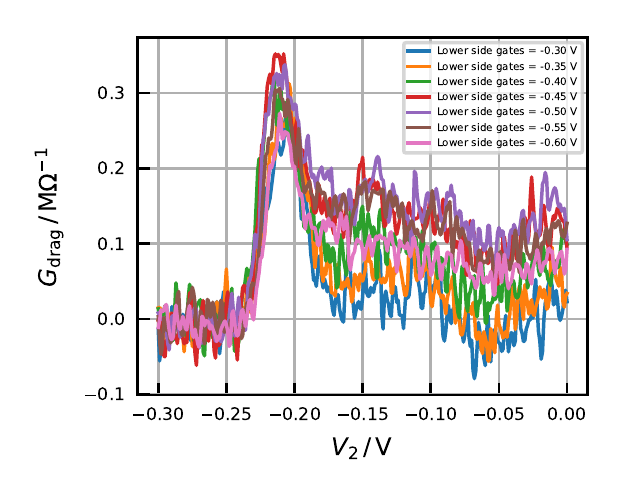}}
	\caption{Differential drag conductance as a function of the voltage ($V_{\rm 2}$) on gate g2 for different drive-wire-widths. Here we fix $V_{\rm sd}$ to \SI{2}{mV}. To change the width of the drive wire we modified the voltage on all the lower side gates at the same time between \SI{-0.3}{V} and \SI{-0.6}{V}. The correspondence between the colour of the $G_{\rm drag}$ data and the value of the lower side gates is indicated in the figure legend.
	}
	\label{fig:ub2}
\end{figure}
%%%%%%%%%%%%%%%%%%%%%%%%

\newpage
\section{Drag current in the asymmetric drag wire configuration with a potential barrier in the drive wire}
\begin{figure}[htbp]
	\centerline{\includegraphics[width=0.8\textwidth]{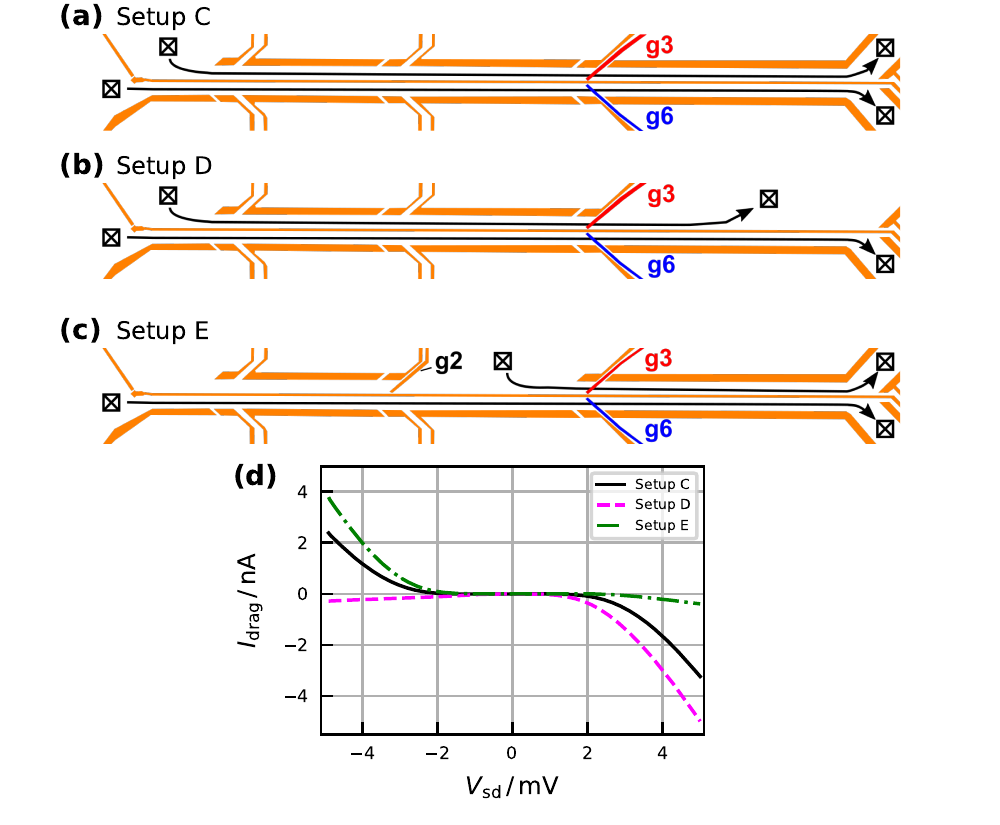}}
	\caption{
		(a - c) Schematic of the device-setup employed in the measurements of (d). Here potential barriers are formed at g3 and g6. The value of the applied voltage is indicated by the dotted circle in Figs.\,\ref{fig:device}c and d in the main paper, respectively. (d) Drag current as a function of $V_{\rm sd}$ observed in the 3 device-setups depicted in (a) - (c).
	}
	\label{fig:lb2}
\end{figure}

%To investigate more this asymmetric phonon-based energy transfer, 
Here we perform the drag current measurement in three different device setups as indicated schematically in Figs.\,\ref{fig:lb2}a-c.
The observed drag currents for the three setups are plotted in Fig.\,\ref{fig:lb2}d.
In setup C we simply form a potential barrier at gates g3 and g6 and set the voltages near the pinch off showing a peak/dip in the differential drag conductance for both gates (the values are also indicated by the dotted circle in Figs.\,\ref{fig:device}c and d in the main paper).
We find a counterflow of electrons (see the black solid curve in  Fig.\,\ref{fig:lb2}d) as for the case when the gates g2 and g5 are used as barriers (see the blue dash-dotted curve in Fig.\,\ref{fig:lb}b in the main paper).
Compared to setup C, in setup D we depolarise the gate on the right of gate g3 which in turn opens the right side of the potential barrier in the drag wire to the reservoir.
This suppresses the 1D channel region of the drag wire to the right of the barrier gate g3.
When $V_{\rm sd}$ is positive and electrons flow from the right to the left in the drive wire, we observe a counterflow drag current (see magenta dashed curve in Fig\,\ref{fig:lb2}d).
On the other hand, when $V_{\rm sd}$ is negative and electrons flow from the left to the right in the drive wire, the drag current is highly suppressed and the counterflow behavior disappears.
This result indicates that the main origin of the observed drag current here is asymmetric about the potential barrier and depends on the asymmetry of the wire and the sign of $V_{\rm sd}$.
The highly suppressed drag current underlines again that the effect of electron excitation is more effective for a 1D quantum wire than for a 2D reservoir.
We assign the small negative drag current for negative $V_{\rm sd}$ to electron excitation by the drive current on the left of the potential barrier as in the case of Fig.\,\ref{fig:ub}c in the main paper.
In setup E, we reverse the situation and reduce the 1D channel region of the drag wire to the left of the barrier gate g3.
To implement this scenario we pinch off gate g2 and depolarise the gate on the left of gate g3 in the drag wire.
In this case, the drag current is suppressed for positive $V_{\rm sd}$ as expected.
These results give further evidence to support the scenario of asymmetric phonon-based energy transfer discussed in Ref.\,\onlinecite{Khrapai2007}.
%%% Figure 4: Drag current for asymmetric device configurations %%%%%%%
%%%%%%%%%%%%%%%%%%%%%%%%

\end{document}